# Physics of Cold Atomic Fermi Gases

E.A. Ayryan [1], K.G. Petrosyan [2],
A.H. Gevorgyan [3], N.Sh. Izmailian [4], K.B. Oganesyan [1,4,*]

[1] Joint Institute for Nuclear Research, Dubna, Russia
[2] Institute of Physics, Academia Scinica, Taipei
[3] Yerevan State University, Yerevan, Armenia
[4] Alikhanyan National Science Lab, Yerevan Physics Institute, Yerevan, Armenia

[*] bsk@yerphi.am

**Abstract**
We consider a cold two-species atomic Fermi gas confined in a trap. We combine the Hermitian coupling between the states (we assume them to be the states with different spins) with the Cooper pairing of atoms with these different spins. This opens up a new prospect for investigation of interplay between various phenomena involving Raman coupling (e.g., atom lasers, dark-state polaritons) and effects caused by Cooper pairing of particles (e.g., superfluidity). We have obtained a threshold of transition from oscillatory to amplifying behavior of matter waves.

## 1. INTRODUCTION

Since recent achievements in producing and manipulating of degenerate Fermi gases [1, 2] there exists a growing interest in fundamental properties of these quantum objects and their applications. Pure samples of fermions in magnetic and optical traps provide with a unique possibility to test a number of quantum statistical physics of loots. That also advances towards realizations of new quantum devices such as quantum computers [3] and intense form ionic beam generators [4, 5].

In this paper we combine two phenomena that were exploited in the physics of cold atoms namely the Raman coupling between different atomic states and the Cooper pairing of particles with different spin states. The Raman coupling played the basic role in construction of bosonic atom lasers [6] like free electron lasers [7-57], or boses, that had been shown to be realizable experimentally [58].

The idea is to couple out the to an untrapped state to get an output coherent beam of atoms. Other bosers proposed are the excitonic [59] and exciton-polariton [60] lasers. Atomic parametric oscillator which produces correlated atomic beams was proposed in [61]. Nowadays we witness a move to implementation and investigation of atomic analogues of optical parametric oscillators [62]. Cooling and thermometry of atomic Fermi gases is reported in recent review [67].

The other phenomenon which we will employ here is the Cooper pairing of particles in trapped Fermi gases [63].

## 2. MODEL HAMILTONIAN AND EQUATIONS

Let us proceed with the elaboration of the interplay of these two phenomena, the Raman coupling and the Cooper pairing of the trapped fermionic atoms. We consider two species of fermions confined in a trap which interact to each other by two-body collisions ($S$ wave scattering).

$$H = \int d^3\mathbf{r} \sum_{\sigma=\uparrow,\downarrow} \Psi_\sigma^\dagger(\mathbf{r}) \left[ -\frac{\hbar^2}{2M}\nabla^2 + V_\sigma(\mathbf{r}) - \mu_\sigma \right] \Psi_\sigma(\mathbf{r})$$
$$+ \varepsilon \left( \Psi_\downarrow^\dagger(\mathbf{r})\Psi_\uparrow(\mathbf{r}) + \Psi_\uparrow^\dagger(\mathbf{r})\Psi_\downarrow(\mathbf{r}) \right) - g\Psi_\downarrow^\dagger(\mathbf{r})\Psi_\uparrow^\dagger(\mathbf{r})\Psi_\uparrow(\mathbf{r})\Psi_\downarrow(\mathbf{r}). \qquad(1)$$

Here, $\Psi_{\uparrow,\downarrow}^\dagger(\mathbf{r})$ (r ) and $\Psi_{\uparrow,\downarrow}(\mathbf{r})$) are the creation and annihilation field operators for the two fermionic species, $V_{\uparrow,\downarrow}(\mathbf{r})$ are the trapping potentials, $\mu_{\uparrow,\downarrow}(\mathbf{r})$ the chemical potentials, $M$ the mass of the particles, and $g$ the interaction parameter. is the parameter responsible for flipping of fermions thus providing the Raman type coupling.
This can be done either by two optical fields or by a r.f. field.
The Hamiltonian (1) describes, e.g., trapped atomic $^6Li$ (fermion) gas under the influence of electromagnetic lield(s) that ensures the Raman transition of atoms from the $|\downarrow\rangle$ trapped state to another trapped state $|\uparrow\rangle$ thus providing with the Raman coupling. The s-wave interaction of the atoms belonging to two different species leads to Cooper pairing among the particles.

To proceed further we will make an approximation assuming that the $|\downarrow\rangle$ and $|\uparrow\rangle$ fermionic states have low densities $\langle \Psi_\sigma^\dagger \Psi_\sigma \rangle << 1$ (we will keep them low throughout the paper). That will lead to the following commutation relations for field operators $\left[ \Psi_\sigma(\mathbf{r}), \Psi_\sigma^\dagger(\mathbf{r}') \right] \square \, \delta^3(\mathbf{r}-\mathbf{r}')$. Then Heisenberg equations of motion for the system are

$$i\hbar\dot\Psi_\uparrow(\mathbf{r}) = \left[ -\frac{\hbar^2}{2M}\nabla^2 + V_\uparrow(\mathbf{r}) - \mu_\uparrow \right] \Psi_\uparrow(\mathbf{r}) + \varepsilon\Psi_\downarrow(\mathbf{r}) - g\,\Psi_\downarrow^\dagger(\mathbf{r})\Psi_\downarrow(\mathbf{r})\Psi_\uparrow(\mathbf{r}),$$
$$i\hbar\dot\Psi_\downarrow(\mathbf{r}) = \left[ -\frac{\hbar^2}{2M}\nabla^2 + V_\downarrow(\mathbf{r}) - \mu_\downarrow \right] \Psi_\uparrow(\mathbf{r}) + \varepsilon\Psi_\uparrow(\mathbf{r}) - g\,\Psi_\uparrow^\dagger(\mathbf{r})\Psi_\uparrow(\mathbf{r})\Psi_\downarrow(\mathbf{r}). \qquad(2)$$

Let us now consider the case of homogenous spatially invariant (uniform) matter fields $\Psi(r,t) = \Psi(t)$. Thus we put $V_\sigma = 0$ inside the trap and $V_\sigma = \infty$ on the boundary.
The equations become

$$i\hbar\dot\Psi_\uparrow = -\mu_\uparrow \Psi_\uparrow(\mathbf{r}) + \varepsilon\Psi_\downarrow(\mathbf{r}) - g\,\Psi_\downarrow^\dagger \Psi_\downarrow \Psi_\uparrow,$$
$$i\hbar\dot\Psi_\downarrow = -\mu_\downarrow \Psi_\uparrow(\mathbf{r}) + \varepsilon\Psi_\uparrow(\mathbf{r}) - g\,\Psi_\uparrow^\dagger \Psi_\uparrow \Psi_\downarrow. \qquad(3)$$

with the attractive interaction between particles due to collisions.

## 3. MEAN-FIELD THEORY FOR MATTER WAVES

Let us now assume that Cooper pairing takes place and introduce the gap $\langle \Psi_\downarrow \Psi_\uparrow \rangle = \Delta$ assuming it to be constant (time-independent). We also assume the thermodynamical equilibrium between the species $\mu_\downarrow = \mu_\uparrow = \mu$. Now in the interaction picture ($\Psi_\sigma e^{i\mu t/\hbar} \to \Psi_\sigma$) the equations become

$$i\hbar \dot{\Psi}_\uparrow = \varepsilon \Psi_\downarrow(\mathbf{r}) - g\Delta \Psi_\downarrow^\dagger,$$
$$i\hbar \dot{\Psi}_\downarrow = \varepsilon \Psi_\uparrow(\mathbf{r}) - g\Psi_\uparrow^\dagger. \tag{4}$$

Using these equations we obtain the following equation for the matter fields

$$\ddot{\Psi}_\sigma + \frac{\varepsilon^2 - g^2\Delta^2}{\hbar^2} \Psi_\sigma = 0 \tag{5}$$

for both spin components.
Let us summarize what we have done so far. We started with a rather general interaction between atoms belonging to two different species with a switching between them via a r.f. field. Then we assumed the atoms to couple into Cooper pairs with the gap being time-independent. All fields arc assumed to be uniform.
These simplifications can be in principle overcome however we will continue with the Eq. (5) that describes matter fields for both spin components.

The Eq. (5) is an equation for a linear oscillator with the effective frequency $\omega^2 = \dfrac{\varepsilon^2 - g^2\Delta^2}{\hbar^2}$.

It is easily seen from it that we have a conventional oscillatory regime in case of $\varepsilon > g\Delta$ and the regime of amplification in the opposite case $\varepsilon < g\Delta$. The inequalities indicate that for weak fields (as well as in case of their absence) there is always an amplification of the matter fields caused by "dissociation" of Cooper pairs.
This is somewhat reminiscent of dissociation of molecules consisting of two fermi atoms [64]. For strong enough fields we have an oscillatory behavior supported by inter-transfer of Cooper pairs into single atoms and backward pairing.

## 4. DYNAMICS IN THE CASE OF TIME-DEPENDENT MEAN-FIELD

Let us now consider the case when the Cooper pairing occurs much faster than the matter fields change. We assume that the characteristic time for single atomic dynamics $\tau_\sigma$ is much bigger than the time $\tau_\Delta$ of Cooper pairs dynamics $\tau_\Delta \ll \tau_\sigma$. In this case we may have the Eq. (4) with time-varying gap function $\Delta(t)$. As an example we can try the soliton profile for the gap [65].

$$\Delta(t) = \frac{\Delta_0}{\cosh \Delta_0 t}. \tag{6}$$

With the assumption of adiabatically changing $\Delta(t)$ we will arrive at the following equation (we also scale time ($t/\hbar \to t$)

$$\ddot{\Psi}_\sigma + \left[\varepsilon^2 - \frac{g^2 \Delta_0^2}{\cosh^2 \Delta_0 t}\right] \Psi_\sigma = 0 \tag{7}$$

This is a Schrödinger-like equation with an effective potential with a soliton shape. Notice that the $\frac{g^2\Delta_0^2}{\cosh^2\Delta_0 t}$ term goes to zero with time. It means that even if at the beginning we had $\varepsilon < g\Delta_0$ amplification condition with time the system turns to oscillations.

This helps to avoid infinite growing of matter fields thus always keeping $\langle \Psi_\sigma^\dagger \Psi_\sigma \rangle$ finite. We also point out that the Eqs. (4) could be solved directly without the adiabatic assumption for the gap function.

## 5. FLUCTUATING GAP FUNCTION

Let us now consider the case of the fluctuating gap function that changes in time as follows,

$$\Delta(t)^2 = \Delta_0^2 + \sigma\xi(t); \quad \langle \xi(t) \rangle = 0, \tag{8}$$

where $\xi$ is a random function with zero average. Thou we can rewrite the Eq. (5) in the following form,

$$\ddot{\Psi}_\sigma + \omega_0^2[1 + \alpha\xi(t)]\Psi_\sigma = 0, \tag{9}$$

where

$$\omega_0^2 = \varepsilon^2 - g^2\Delta_0^2. \tag{10}$$

For the averaged quantity $\Psi_\sigma(t)$ we obtain the following equation [66]

$$\langle \ddot{\Psi}_\sigma \rangle + \frac{1}{2}\alpha^2\omega_0^2 c_2 \langle \dot{\Psi}_\sigma \rangle + \omega_0^2\left(1 - \frac{1}{2}\alpha^2\omega_0 c_1\right)\langle \Psi_\sigma \rangle = 0, \tag{11}$$

$$c_1 = \int_0^\infty \langle \xi(t)\xi(t-\tau) \rangle \sin 2\omega_0\tau \, d\tau, \tag{12}$$

$$c_2 = \int_0^\infty \langle \xi(t)\xi(t-\tau) \rangle (1 - \cos 2\omega_0\tau) d\tau. \tag{13}$$

Notice that for white noise $\langle \xi(t)\xi(t-\tau) \rangle = \delta(t)$ and therefore $c_{1,2} = 0$. However if this is not the case we have a damping. We then can consider cases of underdamped oscillations (neglect dissipation term) and over-damped dynamics (neglect the term with the second lime derivative).

## ACKNOWLEDGMENTS

We thank R. Onofrio for paper [67].